# STATISTICAL ANALYSIS OF SOFT X-RAY SOLAR FLARES DURING SOLAR CYCLES 21, 22 AND 23

Navin Chandra Joshi*[, a], Neeraj Singh Bankoti[a], Seema Pande[b], Bimal Pande[a], Wahab Uddin[c] and Kavita Pandey[a]

[a]Department of Physics, DSB Campus, Kumaun University, Naini Tal – 263 002, Uttarakhand, India

[b]Department of Physics, MBPG. College, Haldwani, Kumaun University, Naini Tal, Uttarakhand, India

[c]Aryabhatta Research Institute of Observational Sciences, Naini Tal – 263 129, Uttarakhand, India

*E-mail address: njoshi98@gmail.com*

## ABSTRACT

This paper presents statistical analysis of Soft X-ray (SXR) flares during the period January 1976 to December 2007 covering solar cycles (SCs) 21, 22, and 23. We have analysed north-south (N-S) and east-west (E-W) asymmetry of SXR at total (1° - 90°), low (1° - 40°) and high (50° - 90°) latitudes and center meridian distances (CMDs). We have also presented the N-S and E-W asymmetry of different intensity classes (B, C, M, and X) during the period of investigation. A slight southern and eastern excess is found after analysis during SCs 21, 22, and 23. N-S distribution shows that the SXR flare events are most prolific in the 11-20° latitude band in the northern and southern hemispheres whereas E-W distribution does not show any prolific band. We found that the annual N-S and E-W hemispheric asymmetry at low latitudes and CMDs is the same as total latitudes and CMDs respectively. E-W asymmetry is different at low and high CMDs. We found that the N-S and E-W asymmetry which often peaks near the activity minimum is in agreement with the theoretical dynamo models. Our results show that N-S asymmetry is statistically more significant than E-W asymmetry. It is revealed that the SXR flare activity (M and C class flares) during SC 23 is low as compared to the SC 22, indicating the violation of Gnevyshev-Ohl rule. The B class flare activity is higher for SC 23 whereas C, M and X class activities are higher for SC 21. We have also analysed the flare evolution parameters, i.e. duration, rise time, decay time and event asymmetry for total SXR as well as for different classes during last three SCs. The duration, rise time and decay time increase with increasing intensity class. The increase is more pronounced for the duration and decay time than for the rise time for SCs 21 and 22 while for SC 23 it is the same for duration, rise time and decay time. On analyzing event asymmetry indices, we found more positive values during SC 21 (≈ 64.86%) and SC 22 (≈54.31%), but for SC 23 we have more negative values (≈48.08%). Our study shows that during SC 23 we have more SXR flare events having shorter decay time than the rise time as compared to SCs 21 and 22.

## 1. INTRODUCTION

The Sun is the only star whose magnetic activity can be observed and studied in detail. It exhibits 11 and 22 year spot and magnetic cycles and a number of periodicities which are explained by an oscillatory magnetic dynamo and other mechanisms. The distribution of sunspots and other activity phenomena in the solar photosphere reflects the magnetic field structure in the convection zone and provides strong observational constraints on the solar dynamo theory. Being the nearest star it provides a unique opportunity to study the various ongoing physical processes and magnetic activity in other stars in spatial and temporal resolutions.

Understanding the nature and characteristic of the solar activity cycle and heliospheric asymmetries has been a major aspect for solar astronomy. 11-year period of the solar cycle (SC) was discovered about 150 years ago (Schwabe, 1844). A drift of active latitudes towards equator and a variation of the rotation rate of the Sun with latitude were noted in sunspot motions shortly thereafter (Carrington, 1858, 1859). The statistical investigations of the characteristics of solar flares started in 1930s, when a worldwide surveillance of the Sun based on Hale's spectrohelioscope was established (Cliver, 1995). Since then many papers have been published (Garcia, 1990; Mariş et al., 2002; Temmer et al., 2001; Veronig et al., 2002) which deal with the analysis of different statistical aspects of solar flares. Statistical studies of hemispheric asymmetries present a better understanding of both observational and theoretical character of SCs.

Solar activity phenomena are not uniformly distributed on solar hemispheres (northern, southern, eastern and western). The north-south (N-S) and east-west (E-W) distribution and asymmetries of several manifestations of solar activities such as flares, magnetic flux, sunspot numbers, sunspot area and solar active prominences (SAP) have been studied by various authors (Bell, 1962; Garcia, 1990; Howard, 1974; Hansen and Hansen, 1975; Joshi and Joshi, 2004; Joshi and Pant, 2005; Joshi et al., 2009; Li et al., 1998; Li et al., 2003; Li et al., 2009; Mariş et al., 2002; Oliver and Ballester, 1994; Reid, 1968; Roy, 1977; Swinson et al., 1986; Temmer et al., 2001; Uddin et al., 1991; Verma, 1987a, 1987b; Verma, 2000). Gnevyshev and Ohl (1948) gave an empirical rule (G – O rule) which states that the sum of sunspot numbers over an odd cycle exceeds that of the preceding even cycle. Bell (1962) found a long-term asymmetry in the sunspot area data for SCs 8 through 18. Howard, (1974) examined the N-S distribution of solar magnetic flux for the period 1967-1973 and found that about 95% of the total magnetic flux of the Sun is confined to latitudes below 40° in both the hemispheres. Some authors (Ataç and Özgüç, 1996; Li et al., 2002; Vizoso and Ballester, 1990) used the straight line fit to the yearly values of the N-S asymmetry indices of different solar activity features during a long period, and found that for each of the four cycles, the slope of the straight regression line changes sign, suggesting a periodic behavior. Garcia (1990) studied the N-S asymmetry of soft X-ray (SXR) flares (class $\geq$ M1) during the SCs 20 and 21 and found that the preponderance of flares occurs in the northern hemisphere during the early part of the cycle and moves towards the southern hemisphere as the cycle progresses. Li et al. (1998) studied the N-S asymmetry of SXR flares (M $\geq$ 1) during the maximum period of SC 22 and found that during this cycle, asymmetry favoured the southern hemisphere. Li et al. (2003) studied the N-S asymmetry of the solar active prominence (SAP) at low ($\leq$ 40°) and high ($\geq$ 50°) latitudes from 1957-1998 and found a little connection between asymmetry at low and high latitudes. Recently Joshi et al. (2009) presented N-S distribution and asymmetry of different limb and disk features of SAP as well as total SAP data during cycle 23 and obtained that during the rising phase of the cycle the number of SAP events are roughly equal in the northern and southern hemispheres and activity dominates southern hemisphere after 1999. They also discussed a comparison among the last four SCs (i.e., 20, 21, 22 and 23) and found that the SAP activity during the cycle 23 is low as compared to previous four SCs. All these studies of long term behavior of different flare activities, N-S asymmetry and distribution play a key role in understanding the dynamics of SCs and the underlying dynamo process.

In comparison there are relatively few studies pertaining to the longitudinal (E-W) distribution and asymmetry of solar activity (Heras et al., 1990; Joshi et al., 2009; Letfus, 1960; Letfus and Růžičkoυá-Topoloυá, 1980; Skirgiello, 2005; Temmer et al., 2001). Letfus and Růžičkoυá-Topoloυá (1980) have analysed the E-W asymmetries of Hα flares from 1959 to 1976 and concluded that these are statistically significant only for certain periods of time. Li et al. (1998) found that the E-W asymmetry was not significant in the case of SXR flares (class ≥ M1) during the maximum phase of SC 22. Temmer et al. (2001), found evidence for the existence of a small but significant E-W asymmetry in the occurrence of Hα flares. Verma (2000) also examined E-W asymmetry of SAP for the period 1975-1998 and did not report any significant asymmetry. Joshi and Pant (2005) investigated the E-W asymmetry of solar Hα flares and Joshi et al. (2009) the SAP during SC 23 and reported a small E-W asymmetry. Moreover, E-W asymmetry is a controversial issue in the sense that it is dependent on the observer's position and no obvious physical reason has been worked out till date for it to exist over a long period (Heras et al., 1990).

Along with the observational work, theoretical modeling is an additive tool to understand the variation of SC and Heliospheric asymmetries of solar activity. These asymmetries may yield important detailed information for the nature of solar dynamo action. Hence more attention has been given to the study of N-S asymmetries of different solar activity manifestations and various types of dynamo models have been worked out in the recent past (Chatterjee et al., 2004; Goel and Choudhuri, 2009; Nandy and Choudhuri, 2002; Ossendrijver et al., 1996; Pulkkinen et al., 1999). Ossendrijver et al. (1996) presented a simple axisymmetric mean field dynamo model for the Sun and concluded that the stochastic fluctuations most probably originate in the giant convective cells. They also discussed various aspects of N-S asymmetry on the basis of their dynamo model. Recently, Goel and Choudhuri (2009) have analysed and simulated asymmetries in SCs during the twentieth century using their own solar dynamo model. They have found a good agreement between observation and theory with correlation coefficients 0.73.

Statistical investigations on temporal aspects of Hα and SXR solar flares have been investigated in various studies in the past (Barlas and Altas, 1992; Culhane and Phillips, 1970; Drake, 1971; Pearce and Harrison, 1988; Reid, 1968; Temmer et al., 2001; Veronig et al., 2002). Temmer et al. (2001) statistically analysed the temporal behavior as well as the spatial distribution on the solar disk of Hα flares from January 1975 to December 1999. Veronig et al. (2002) presented a statistical analysis of SXR flares pertaining to their temporal properties i.e., duration, rise time and decay time during the period 1976-2000. Study of temporal aspects and event asymmetry of Hα and SXR flares and their relation with the importance and intensity classes respectively is crucial to the study of different phases of flares i.e. the rising (heating-up of the chromospheric plasma), the maximum, and the decay (cooling) phase.

In this paper, we have analysed the N-S and E-W distribution and asymmetry of SXR flare events separately at total (1° - 90°), low (1° - 40°) and high (50° - 90°) latitudes as well as that of different SXR classes (B, C, M and X) for the last three SCs, i.e., 21, 22 and 23. The temporal aspects and event asymmetry for the same period have also been studied. In Section 2, data set is described. Statistical analysis methods are presented in Section 3 and Section 4 contains a brief description of distribution wherein N-S asymmetry (Section 4.1), E-W asymmetry (section 4.2), temporal aspects and event asymmetry (Section 4.3) of SXR flares have been described. Section 5 deals with discussion and conclusions.

## 2. DATA

For the present analysis data for the period of January 01, 1976 to December 31, 2007 have been downloaded from national geophysical data center (NGDC) anonymous ftp server: *ftp://ftp.ngdc.noaa.gov/STP/SOLAR_DATA/SOLAR_FLARES/XRAY_FLARES*. During this period the occurrence of 63594 SXR flares are reported. To perform the study, we have split the data in three parts corresponding to SC 21 (January 1976 to December 1985) with 3652 days data, SC 22 (January 1986 to December 1995) with 3652 days data and SC 23 (January 1996 to December 2007) with 4383 days data. In NGDC data there are a number of events for which heliographic latitude and CMD are not given. Also in the downloaded data there are events which lie on zero latitude and CMD. After neglecting these events, 34377 events for N-S and 34247 events for E-W, were obtained for distribution and asymmetry analysis. In our data base we have 99.64% of total SXR flare events at low latitudes and only 0.14% SXR events at high latitudes. We have 51.77%, 36.12% of total SXR events at low and high CMDs respectively. The original data are listed in Table 1 for SCs 21, 22, and 23. The present analysis also deals with temporal aspects of SXR flare events. For this we have rejected the events for which the listed time is marked as inaccurate. Thus we get a total of 55431 events for temporal parameter and event asymmetry analysis.

Table 1 lists the total number of SXR flares reported for the period under investigation, subdivided into different SXR flares classes (B, C, M, and X). Since class A events are very few in number, we have combined these events into class B. It can be seen form this table that the bulk of flares belong to class C for all the three SCs and class X events are much less in number. The percentage of C, M, and X class events during SC 21 is large as compared to SCs 22, and 23. During SC 23 the number of class M and X flare events are small as compared to SCs 21 and 22. SC 23 produced maximum number of class B flares (36.56%) and total SXR flare events. For the three SCs we list the average monthly rate of SXR flares for total as well as different class data in Table 2. It is clear from this table that SC 21 exhibit higher average monthly rate of class C, M and X flare events whereas SC 23 shows a higher rate of class B flare events. The flare activity in terms of class B and C flare events during SC 23 is high compared to SC 22. Class M and X flare activities during SC 23 are low in comparison to SCs 21 and 22. In terms of total number of SXR flare events SC 23 depicts higher average monthly flare rate than SCs 21 and 22. Fig. 1 represents the plot for total number as well as for different importance classes of flares. It can be clearly seen that the variation of class B flares is different as compared to the other classes and the total SXR flares.

## 3. STATISTICAL METHODS

We have calculated the N-S ( $A_{NS}$ ) and E-W ( $A_{EW}$ ) asymmetry indices by using the formula

$$A_{NS} = \frac{N-S}{N+S} \text{ and } A_{EW} = \frac{E-W}{E+W} \text{ respectively.} \quad (1)$$

Here $N$ and $S$ are the numbers of SXR flare events observed in the northern and southern halves of the solar disk. If $A_{NS} > 0$, the activity in the northern hemisphere dominates or else it will dominates the southern hemisphere. The E-W asymmetry index is defined analogously. To study the statistical significance of the N-S and E-W asymmetry indices we have used binomial probability distribution. Let us consider n objects in 2 classes. The binomial formula to compute the actual probability $p(k)$ of obtaining $k$ objects in class 1 and $(n-k)$ objects in class 2 is given by (Li and Gu, 2000; Vizoso and Ballester, 1990).

$$P(k) = \frac{n!}{(n-k)!k!}\frac{1}{2^n}. \quad (2)$$

The probability of obtaining more than $d$ objects in class 1 is

$$P(\geq d) = \sum_{k=d}^{n} p(k) \quad (3)$$

In general, when $P(\geq d) > 10\%$ implies a statistically insignificant result (flare activity should be regarded as being equivalent for the two hemispheres); when 5% < $P(\geq d)$ < 10% it is marginally significant; when $P(\geq d) < 5\%$ and $P(\geq d) < 1\%$ we have statistically significant and highly significant results respectively (flare occurrence in not due to random fluctuations) (Li et al., 2001; Li et al., 2003; Oliver and Ballester, 1994).

For the study of temporal behavior of SXR flares we have used some statistical parameter described below. Since the distributions of duration, rise time and decay time are asymmetric, they are better represented by the median ( $\bar{x}$ ) than the arithmetic mean. To know the measure of dispersion we have applied the median absolute deviation ( $\bar{D}$ ), which can be calculated as

$$\bar{D} = Median\left\{\left|x_i - \bar{x}\right|\right\}, \quad (4)$$

where $x_i$ denote the data values and $\bar{x}$ is the median of the $x_i$. As a measure of statistical significance we make use of the 95% confidence interval ( $c_{95}$ ), $\bar{x} \pm c_{95}$ with

$$c_{95} = \frac{1.58\,(Q_3 - Q_1)}{\sqrt{n}}, \quad (5)$$

$Q_1$ and $Q_3$ denote the first and the third quartile, respectively, n being the total number of data values (Veronig et al., 2002). We have also calculated the $90^{th}$ percentile ( $P_{90}$ ), which states that only 10 % of the events have a value larger than $P_{90}$. To study a characterization of the degree of asymmetry of a distribution around its mean we have also calculated the skewness of the distribution. For distributions with positive skewness the median value is smaller than the arithmetic mean, whereas for distributions with negative skewness it is larger than arithmetic mean.

We computed the event asymmetry index ($A_{ew}$) in order to characterize the proportion of the rise time and the decay time of a flare event defined as

$$A_{ew} = \frac{t_{decay} - t_{rise}}{t_{decay} + t_{rise}}, \quad (6)$$

where $t_{rise}$ is the rise time, and $t_{decay}$ the decay time. The event asymmetry index is a dimension-less quantity. A value close to zero states that the rise and the decay times are roughly equal. And if $A_{ev} > 0$ the decay phase is longer than the rising phase otherwise the rising phase will be longer than the decay phase (Temmer et al., 2001).

## 4. ANALYSIS AND RESULTS

### 4.1 NORTH-SOUTH ASYMMETRY OF SXR FLARES AT TOTAL, LOW AND HIGH LATITUDES DURING SOLAR CYCLES 21, 22 AND 23

To investigate the existence of a spatial distribution of flares with respect to heliographic latitudes, we have evaluated the total number of flares in the interval of 10° latitude for northern and southern hemispheres for SCs 21, 22 and 23 in Table 3. Fig. 2 represents the yearly number of SXR flares at low and high latitudes. Fig. 2 shows that both the SXR flare events at high and low latitudes are not uniformly distributed in either the northern or the southern hemispheres. In Fig. 3 we have plotted the heliographic latitudes verses number of total SXR flares and different classes for SCs 21, 22 and 23. All curves show a pronounced peak at 11-20° latitude band on both sides of solar equator. Here 0° represents the equator of the Sun. The N-S asymmetry indices for total SXR flares and different intensity classes (B, C, M, and X) based on annual counts from 1976 to 2007 have been plotted in Fig. 4. Out of 32 N-S asymmetry indices, 21 values come out to be highly significant, 4 values come out to be significant. Out of 32 N-S asymmetry indices 2 come out to be marginally significant and 5 are insignificant. It is clear from this figure that the variation of N-S asymmetry for total and B, C, M, and X flare events is more or less similar. Shift of asymmetry from one hemisphere to the other is studied by fitting a straight line to the yearly values of the asymmetry indices for total SXR as well as for different classes during SCs 21, 22 and 23. These fitted straight regression lines are represented in Fig. 5. The plots clearly show that the slopes of the class X events are quite different from the other classes and total SXR flare events for SCs 22 and 23. In Fig. 6 we have plotted the N-S asymmetry indices versus years separately at low and high latitudes. At low latitude, out of 32 N-S asymmetry values 21 come out to be highly significant, 4 come out to be significant and 2 and 5 come out to be marginally significant and insignificant respectively. At high latitude 2 out of 32 N-S asymmetry values come out to be significant and rest of the values are insignificant. Highly significant, marginally significant, significant and insignificant values of asymmetry indices are marked with different symbols in Fig. 4 and Fig. 6. In Fig. 7 we have fitted a straight line for the yearly values of the asymmetry of the SXR flare events at total, low and high latitudes for SCs 21, 22 and 23 respectively.

### 4.2 EAST-WEST ASYMMETRY OF SXR FLARES AT TOTAL, LOW AND HIGH CMDs DURING SOLAR CYCLES 21, 22 AND 23

Table 4 presents the total number of flares at longitudinal intervals of 10° from the CMD towards the east and west limbs for SCs 21, 22 and 23. Fig. 8 represents the yearly number of SXR flares at high and low CMDs. It is clear from this figure that the SXR flare events at high and low CMDs are not uniformly distributed in the eastern and western hemispheres. In Fig. 9 we have plotted the heliographic CMDs verses number of SXR flares and different classes for SCs 21, 22 and 23. It is clear from this figure that there is no pronounced peak obtained. The E-W asymmetry indices for total SXR flares based on annual counts from 1976 to 2007 have been plotted in Fig. 10. Also plotted in the figure are the E-W asymmetry indices of different intensity classes (B, C, M, and X) for the same period. Out of 32 E-W asymmetry indices values, 15 values turn out to be highly significant and 4 come out to be significant. Out of 32 E-W asymmetry values 1 and 12 values come out to be marginally significant and insignificant respectively. In Fig. 11 we have fitted a straight line to the yearly values of E-W asymmetry indices for total SXR data as well as for different classes for SCs 21, 22 and 23. Form this figure we can find out that for SC 21 the slopes of the fit for total SXR and for different intensity classes are different whereas for SCs 22 and 23 the slopes are the same for all classes and total SXR events. In Fig. 12 we have plotted the E-W asymmetry indices versus years separately at low and high CMDs. At low CMDs, out of 32 E-W asymmetry values 7 turn out to be highly significant, 6 significant and 4 and 15 values turn out to be marginally significant and insignificant respectively. At high CMDs, out of 32 E-W asymmetry indices values 12 turn out to be highly significant, 6 significant and 2 and 12 values turn out to be marginally significant and insignificant respectively. At high CMDs E-W asymmetry is statistically more significant than at low CMDs. Highly significant, marginally significant, significant and insignificant values of asymmetry indices are marked with different symbols in Fig. 10 and Fig. 12. In Fig. 13 we have fitted a straight line to the yearly values of the E-W asymmetry of the SXR events at total, low and high CMDs for SCs 21, 22 and 23 respectively and show that for SCs 21 and 22 the slopes at high and low CMDs are different whereas for SC 23 it is the same.

### 4.3 TEMPORAL ASPECTS AND EVENT ASYMMETRY

In Table 5 we give a list of various statistical measures characterizing the duration, rise time and decay time of the SXR data, namely the mean, the median, the mode, and the $P_{90}$ for SCs 21, 22, and 23. In Table 6 we list the median values (plus $c_{95}$), $\bar{D}$ and the $P_{90}$ values of the temporal parameters calculated for different classes of SXR flares as well as for total SXR flares for SCs 21, 22 and 23. It is clear form this table that all the temporal parameters i.e., duration, rise time and decay time increase with the flare intensity class (from class B to class X). The differences from one class to the other are larger than the $c_{95}$, indicating the statistical significance of the effect. From this table we have found that for SCs 21 and 22 the increase is more pronounced for duration (by a factor between 6-7) and decay time (factor 9-10) than the rise time (factor 3) but for SC 23 this increase is by a factor between 2-3 for duration, rise time and decay time. In Table 7 we list the median values with $c_{95}$, 10$^{th}$ percentile ($P_{10}$) and $\bar{D}$ values of the event asymmetries for different intensity classes as well as for total SXR data for the last three SCs. It can be seen that the asymmetries increase with increasing intensity class for SCs 21 and 22 but for SC 23 it is decreasing with increasing intensity classes. Since the difference of event asymmetries between various classes are larger than the $c_{95}$, the effect can be considered as statistically significant. Fig. 14 shows the distributions of the event asymmetries calculated versus number of flares for SCs 21, 22 and 23. For SC 21 distribution of event asymmetry reveals a negative skewness, whereas for SCs 22 and 23 distributions show a positive skewness. For SCs 21 and 22 median values of event asymmetries are ≈ 0.3 and ≈ 0.1 respectively, which imply that for 50% of events the decay phase is larger by 1.85 and 1.22 times respectively than the rising phase. Whereas for SC 23 median value of asymmetry is ≈ 0.0, which means that for 50% of events the decay times are equal or greater to the rise times. For SCs 21, 22 and 23 the values of $P_{10}$ are ≈ - 0.33 which means we have 90% of events having event asymmetry indices greater than -0.33.

## 5. DISCUSSION AND CONCLUSIONS

The SXR flare data in the period 1976 to 2007 are used to study the N-S and E-W distribution and asymmetry of total and also of different intensity classes at total (1° - 90°), low (1° - 40°) and high (50° - 90°) latitudes and CMDs during the SCs 21, 22 and 23. The data is also used to study the temporal aspects and event asymmetry of SXR flares for these three SCs. The results obtained are the following:

1. From the present study it is concluded that low latitudes (1° - 40°) are the active latitudes and produce maximum number of SXR flares.

2. N-S distribution study of total SXR flare events and different importance class events show similar distribution whereas E-W distribution study shows same variation for total SXR and class C flare events for SCs 21, 22 and 23. For N-S distribution the flares are most prolific between 11° to 20° latitudes whereas for E-W distribution no pronounced peak is obtained.

3. The annual variation of N-S and E-W asymmetry at low latitudes and CMDs is similar to that of total latitudes and CMDs respectively. E-W asymmetry is different at low and high CMDs for SCs 21 and 22 but it is similar at low and high CMDs for SC 23.

4. From Table 2 and 3 it can be seen that during SCs 21, 22 and 23 the activity was southern and eastern hemisphere dominated.

5. Our statistical study shows that N-S asymmetry is more statistically significant than E-W asymmetry.

6. The duration, rise time and the decay time increases with increasing intensity class. For SCs 21 and 22 this phenomenon is more pronounced for the duration and decay time as compared to rise time, but for SC 23 it is the same for duration, rise time and decay time.

7. From event asymmetry analysis we found more positive values of event asymmetry indices during SC 21 (≈ 64.86%) and SC 22 (≈ 54.31%) compared to SC 23, for which we found more negative values (≈ 48.08%).

From Fig. 1 it is evident that during the minimum activity of SCs the class B events dominate whereas for the maximum activity other classes dominate. This is because during the period of maximum activity the X-ray background is too high to detect class A and B flares form full-disk measurements. The increased X-ray background during maximum solar activity may be due to emission from many flare events and also due to a steady coronal heating mechanism (Feldman et al., 1997).

Garcia (1990) gave the distribution of M ≥ 1 class flares from 1969 to 1988 with respect to the heliographic latitude and found that the majority of the flare events occurred within ± 30° latitudes. Li et al. (1998) examined the latitudinal distribution of flares during the maximum period of SC 22 and found that the majority of flares occurred in the latitudes between 8° - 35° in both the hemispheres. Recently Joshi et al. (2009) have reported that during SCs 20, 21, 22 and 23 a larger number of SAP also lie in the latitude band between 1° - 40° than that of high latitudes. In our study most of the SXR events (99.35%) occurred at low latitudes (1° - 40°), which is the same as found in above studies.

Verma et al. (1987b) and Uddin et al. (1991) studied the N-S distribution of major flares during SCs 19 and 20 and sunspots for the period 1967-1987 (SCs 20 and 21) respectively and found that the ±11-20° latitude belt is most prolific for the occurrence of major flares and various spot types irrespective of magnetic-field range. Joshi and pant (2005) and Verma (2000) presented the distribution of Hα flares and SAP and found that in both the hemispheres (N and S) the flares and SAP are more prolific between 11° to 20° latitudes. In our study we have also found the same result for SXR flares and its different intensity classes (Fig. 3). E-W distribution is not found prolific in any CMD bands (Fig. 9).

Li et al. (2003) presented the asymmetry of SAP at low and high latitudes from 1957 through 1998 and found that the annual hemispheric asymmetry exists at low latitudes, but strangely, a similar asymmetry does not seem to occur for SAPs at high latitudes. Recently Li. et al. (2009) studied the N-S asymmetry during SC 23 using sunspot group and sunspot area and found out that the asymmetry signs of solar activity at both the low (>0° - <10°) and high (≥ 25° - ≤ 90°) latitudinal bands are mostly the same but at times different for the middle latitudinal band (≥10° - <25°). In our study we have found that N-S asymmetry at low latitudes is of similar nature to that of total latitudes for all three SCs (Fig. 6 and 7). E-W asymmetry for SCs 21 and 22 (Fig 12 and 13) is different at low and high CMDs whereas during SC 23 it has the same nature. At low CMDs it is shifting from west to east whereas at high CMDs it is shifting in the reverse directions for SCs 21 and 22 but for SC 23 this remains the same at total, low and high CMDs surface (Fig. 13).

Temmer et al. (2001), Joshi and Joshi (2004) and Joshi et al. (2009) studied N-S asymmetry during SCs 21, 22 and 23 by taking Hα solar flare, SXR flare index and SAP respectively and found a significant N-S asymmetry with a prolonged southern excess giving similar results as reported in our study. Vizoso and Ballester (1990), Ataç and Özgüç (1996) and Li et al. (2002) found that the slope of the straight regression line changes every four cycles, which could suggest some kind of periodic behavior in the N-S asymmetry by which the activity in one hemisphere is more prominent during the ascending phase of the cycle whereas during the descending phase the activity becomes more pronounced in the opposite hemisphere. Similar behavior has been obtained in our study for SCs 21, 22 and 23 (Fig. 7) because the activity in the northern hemisphere is more dominant during the ascending phase of cycle but during the descending phase the activity becomes dominant in the southern hemisphere. Joshi and Joshi (2004) have also reported a similar behavior by fitted straight line to asymmetry time series for cycles 21, 22 and 23 using SXR flare index. N-S asymmetry can be explained in terms of 'superactive regions' and 'active zones'. Superactive regions are large, complex, active regions containing sunspots and produce majority of solar flares which appear frequently in certain areas of the Sun, the so-called 'active zones' (Bai, 1987, 1988). The N-S asymmetry therefore can be attributed to the existence of active zones in the northern and southern hemispheres, which can persist over longer periods of time. According to Verma (1993) the reason for the N-S asymmetry period is not known, but perhaps it may be due to the asymmetry in the internal magnetic structure of the Sun. Tritakis et al. (1997) predicted the existence of a time difference in the development of solar activity in the northern and southern hemispheres and this may be a possible explanation for N-S asymmetry. The magnetic cycle of the Sun is believed to be produced by a flux transport dynamo operating in the Sun's convection zone (Choudhuri, 2003; Guerrero and Muňoz, 2004; Nandy and Choudhuri, 2002; Wang et al., 1991). The Babcock-Leighton process of poloidal field generation is believed to be the main source of irregularity in SCs. N-S asymmetry of solar activity of a SC should tend to get reduced as the cycle progresses, and the hemispheric asymmetries should be expected to continuously get washed away until the randomness in the Babcock-Leighton process creates fresh asymmetries towards the end of the cycle (Choudhuri et al., 2007; Goel and Choudhuri, 2009). These discussions lead to our conclusions that N-S asymmetry is a real phenomenon and is not due to stochastic or random fluctuations.

Slight but significant E-W asymmetries with a prolonged eastern excess have been observed for SCs 21 and 22 (Temmer et al., 2001; Verma, 2000) whereas for SC 23 there is a small but prolonged western excess (Joshi and Pant, 2005; Joshi et al., 2009). In our study we have found small but prolonged eastern excess for all three SCs 21, 22 and 23. Li et al. (1998) also studied E-W asymmetry and reported an insignificant but a non–uniform flare distribution in CMDs. Heras et al. (1990) analysed the E-W solar flare distribution from 1976-1985 and found a pronounced and prolonged E-W asymmetry in flares and subflares. They also concluded that simple random distribution of flares over the solar disk cannot account for the asymmetries found, but they can be explained in terms of the transit of active regions in front of the observer's position. Recently Joshi et al., (2009) have also reported a small but significant E-W asymmetry of SAP from 1963 to 2007. In the past a number of studies have been performed to study the E-W asymmetry of different solar activity features however, E-W asymmetry till date is unexplained and remains a controversial issue.

From Tables 1 and 2 it is clear that the flare activity of SXR (class M and X flares) during SC 23 is low compared to SC 22. This clearly indicates the violation of Gnevyshev-Ohl (G-O) rule, for the pair of SCs 22-23, which states that the odd-numbered cycles have greater activity than the preceding even-numbered ones (Gnevyshev and Ohl, 1948). Komitov and Bonev (2001) examined the conditions for violations of the Gnevyshev-Ohl rule, which states that the even-numbered solar cycles have been followed by higher in amplitude odd-numbered ones. They also predicted a high probability for violation of this rule for the pair of the SCs 22-23. Violation of this rule for the same pair has also been pointed out in some studies (Ataç and Özgüç, 2006; Joshi and Pant, 2005).

The shape of N-S and E-W asymmetry curve (Fig. 4 and Fig. 10) shows that the asymmetry has peaked at or around the minimum of solar activity. This result is complementary to the study of N-S asymmetry made by many authors (Ataç and Özgüç, 1996; Visoso and Ballester, 1990). The above result is also in agreement with the result obtained by Ossendrijver et al. (1996) using their theoretical dynamo model of the Sun. We have also found out that N-S as well as E-W asymmetry for SXR and different intensity classes has no relation with a solar maximum year or minimum year during the last three SCs (Fig. 4 and Fig. 10). These results are in agreement with the work done by Verma (2000), who analyzed the variation of N-S and E-W asymmetry of SAP from 1957 to 1998.

Temmer et al. (2001) and Veronig et al. (2002) statistically analysed Hα and SXR flares and found that the temporal parameter increases with increasing importance and intensity classes respectively. Similar result is reported in our analysis for SCs 21, 22 and 23 (Table 6). The results obtained for the mode of duration of SXR flares, 5.0, 7.0 and 7.0 for SCs 21, 22 and 23 (Table 5) respectively, is in agreement with the previous studies. Pearce and Harrison (1988), Feldman et al. (1997) reported a mode in the range 5-10 min. Veronig et al. (2002) found that the average values of the duration, rise time and decay time increase from flare intensity class B to X by a factor of 3. Whereas Temmer et al. (2001) investigated that on the one hand, the increase of the duration with the importance class in particular results from the increase of the decay time, and is significantly more pronounced than the increase of the rise time. On the other hand the event asymmetry also increases with important class. Both the above results show that the cooling phase of the Hα flare is more strongly affected by the flare size than the phase of heating-up the chromospheric plasma at the flare site. In our study we have obtained that for SCs 21 and 22 increase in decay time is more pronounced than the rise time with the intensity class and also the event asymmetry is increasing with intensity classes. But for SC 23 the increase in decay time is the same as the increase in rise time with the intensity class and event asymmetry is decreasing with importance class. On the basis of above discussion we can conclude that for SCs 21 and 22 the decay phase is more affected by the peak burst intensity of SXR flares than the rising phase. However, SC 23 does not show this effect. Temmer et al. (2001) also studied the event asymmetry of Hα solar flares from 1975 to 1999 and investigated that there are predominantly positive values of event asymmetry indices. These results are confirmed in our analysis for SCs 21 and 22 but for SC 23 we found more negative values for SXR solar flares.

## ACKNOWLEDGEMENTS

Two of the authors (NCJ and NSB) wish to thank UGC, New Delhi for financial assistance under RFSMS (Research Fellowship in Science for meritorious students) scheme. We gratefully acknowledge the valuable comments and suggestions from the referees which improved the scientific contents of the paper.

**------------**